\documentclass[preprintnumbers,amsmath,amssymb,floatfix,11pt,prd,onecolumn,superscriptaddress,nofootinbib]{revtex4}
\usepackage{latexsym}
\usepackage{epsfig}
\usepackage{amsmath}
\usepackage{epstopdf}
\usepackage{amssymb}
\usepackage{graphicx}
\usepackage{hyperref}
\usepackage{tabularx}
\usepackage{varioref}
\usepackage{placeins}
\newcolumntype{C}[1]{>{\centering\arraybackslash}p{#1}}

\begin{document}
\title{\bf Area Product and Mass Formula for Kerr-Newman Black Hole in Quintessence}
\author{Ayesha Zakria}
\email{ayesha.zakria@bcoew.edu.pk}
\affiliation{Department of Mathematics, Bilquis Post Graduate College for Women, Air University, Islamabad, Pakistan}
\author{Asma Afzal}
\email{asma.afzal7@gmail.com}
\affiliation{Department of Mathematics, Bilquis Post Graduate College for Women, Air University, Islamabad, Pakistan}
\maketitle
\begin{center}
\textbf{\textbf{ABSTRACT}}
\end{center}

In this research work, predominantly we acquire area, angular velocity, entropy, surface gravity and Hawking temperature of inner and outer horizons for Kerr-Newman black hole in presence of quintessence. Additionally, we determine area sum, area product, entropy sum and entropy product. We examine that the area product and entropy product are free from mass $M$ but they surly rely upon the angular momentum $J$, charge $q$, spin parameter $a$ and the normalization factor $c$. We monitor that these thermodynamic products are universal. We investigate that the area sum and entropy sum rely upon the mass $M$, charge $q$, spin parameter $a$ and the normalization factor $c$, so these sums are not universal. The black hole mass and Christodoulou-Ruffini mass for Kerr-Newman black hole in quintessence are also found. We extract the entropy bound from the area bound. We derive the Penrose inequality and discuss the microscopic nature of the entropy.

 \newpage

\section{Introduction}
The area product and entropy product of the multi-horizon black holes has been interrogated in order to interpret the black hole entropy $\delta=$ {\Large $\frac{A}{4}$} at the imperceptible level. Previous researches indicated that the entropy product and area product for multi-horizon stationary, axisymmetric black holes are frequently free from Arnowitt-Deser-Misner (ADM) mass and rely only upon the charge, angular momentum etc. The researchers showed that the enropy of the outer and inner horizons in string theory and $M$-theory community are of the pattern \cite{1}-\cite{3}
 \begin{equation}\label{Z1}
   \delta_{\pm}=2\pi\big(\sqrt{N_{L}}\pm\sqrt{N_{R}}\big),
 \end{equation}
 where $N_{L}$ and $N_{R}$ are named for the left and right moving modes of a weakly coupled $2$-dimensional conformal field theory (CFT). The product $\delta_{+}\delta_{-}=4\pi^{2}\big(N_{L}-N_{R}\big)$ should be represented in integer multiples of $4\pi^{2}$ \cite{2}-\cite{5}. Absolutely, one could set up
 \begin{equation}\label{Z2}
   \delta_{+}\delta_{-}=(2\pi)^{2}(J^{2}+q_{1}q_{2}q_{3}q_{4}),
 \end{equation}
 and
\begin{equation}\label{Z3}
   \delta_{+}\delta_{-}=(2\pi)^{2}(J_{1}J_{2}+q_{1}q_{2}q_{3}),
\end{equation}
for $4$ and $5$ dimensional black holes, respectively. These products are free from mass but surely rely upon charges and angular momenta. In $4$ dimension, the entropy product is represented by one angular momentum $J$ and four charges $q_{1}$, $q_{2}$, $q_{3}$, $q_{4}$. But in $5$ dimension, the entropy product is expressed by two angular momenta $J_{1}$, $J_{2}$ and three charges $q_{1}$, $q_{2}$, $q_{3}$.

The researchers \cite{6}-\cite{9} illustrated that the entropy product of the outer and inner horizons of Kerr-Newman (KN) black hole has the expression
\begin{equation}\label{Z4}
   \delta_{+}\delta_{-}=(2\pi)^{2}\bigg(J^{2}+\frac{q^{4}}{4}\bigg).
\end{equation}
It is really free from the mass $M$ of the black hole. Clearly, it relies upon the angular momentum $J$ and charge $q$ of the black hole.

The universal nature of the entropy product of black ring and black string solutions have been discussed \cite{10}. The essential point is that the entropy product of inner and outer horizons could be adopted to decide whether the Bekenstein-Hawking entropy may be expressed as a Cardy formula, hence supporting some witness for CFT statement of the comparable microstates \cite{10,11,12}.  Apparently, the Curir was the first who worked out at the area sum and entropy sum of Kerr black hole \cite{13}. The researchers have investigated the universal property of entropy sum of the ADS spacetime \cite{14}.

It is a popular reality that outer horizon is an infinite red shift surface although inner horizon is an infinite blue shift surface \cite{15}. Therefore, when a spectator travel across the outer horizon $r = r_{+}$, pursuing a future directed time-like geodesic, is endlessly `lost' to an outer spectator and any radiation sent by such a spectator at the moment of crossing will be infinitely red shifted. Although, when the same spectator travel across the inner horizon $r = r_{-}$, pursuing a future directed time-like geodesic and at the moment of crossing he/she will encounter a panorama of the complete history of the exterior universe and any radiation sent by such a spectator at the moment of crossing will be infinitely blue shifted. This is the elementary dissimilarities between the outer and inner horizons.

The entropy product and area product of multi-horizons are not forever universal i.e., mass freedom and occasionally they also unsuccessful. The author discussed that these possessions are also hold for Kerr-Newman-Taub-NUT (KNTN) black hole \cite{16}. The researcher surveyed that as a result of the existence of NUT (Newman-Unti-Tamburino) parameter what could be the differences are revealed in the Smarr's mass prescription \cite{17,18} and Christodoulou-Ruffini mass prescription counter to KN geometry. He also particularly validated that the first law of black hole thermodynamics does not influence for KNTN black hole. He additionally manifested that Smarr-Gibbs-Duhem relation does not establish for KNTN black hole. In \cite{19}, the authors scrutinized the thermodynamics of Kerr-Newman-Kasuya and Reissner-Nordstr\"{o}m black hole with a global monopole on the outer and inner horizons.

We will discuss the universal behaviour for KN black hole in presence of quintessence \cite{20}. We will illustrate as a result of the existence of normalization factor what would be the differences reveal in the entropy sum, area sum, area product and entropy product in contrast with KN black hole. This is the main goal of this research paper. Applying the entropy product, we also compute the entropy bound, area bound and irreducible mass bound for both the horizons of KN black hole in presence of quintessence. These are fundamentally all geometrical bound which was first  suggested by Penrose \cite{23} and now it is known as Penrose inequality.

The formation of the paper is as follows. In Section \ref{Sec:A2}, we represent that the area product and entropy product of both the outer and inner horizons satisfy the universal properties by cause of the mass independence. Such products imply to be universal in nature. Whereas, the area sum and entropy sum of both the horizons do not show the universal properties by cause of the mass dependence. In Section \ref{Sec:A3}, we prove that the black hole mass or ADM mass can be demonstrated with regard to the area of both horizons. We find the Christodoulou-Ruffini mass prescription for KN spacetime in presence of quintessence in Section \ref{Sec:A4}. We further manifest the Christodoulou's irreducible mass product of the outer and inner horizons are free from mass. Lastly, we conclude in Section \ref{Sec:A5}
\section{Kerr-Newman Black Hole in Quintessence}\label{Sec:A2}
In Boyer-Lindquist coordinates, the KN metric in quintessence is determined by the following parameters i.e., the mass $M$, spin parameter $a$, electric charge $q$, quintessential field parameter $\omega_{q}$, angular momentum $J$ and normalization factor $c$. The metric of charge rotating black hole in presence of quintessence is given by \cite{20}
\begin{eqnarray}\label{1}
ds^{2}&=&-\frac{1}{\Sigma}\big(\Delta-a^{2}\sin^{2}\theta\big)dt^{2}-\frac{2}{\Sigma}\big(r^{2}+a^{2}-\Delta\big)a\sin^{2}\theta dtd\phi+\frac{1}{\Sigma}\sin^{2}\theta\Big(\big(r^{2}+a^{2}\big)^{2}-\Delta a^{2}\sin^{4}\theta\Big) d\phi^{2}\notag\\&&+\frac{\Sigma}{\Delta}dr^{2}+\Sigma d\theta^{2},
\end{eqnarray}
where
\begin{eqnarray}\label{A1}
a&=&\frac{\emph{J}}{\emph{M}},  \notag\\   \Sigma&=& r^{2}+a^{2}cos^{2}\theta, \\    \Delta&=& r^{2}-2Mr+a^{2}+q^{2}-cr^{1-3\omega_{q}}.
\end{eqnarray}
The quintessence field parameter $\omega_{q}$ can take the value between $-1 <w_{q} < -\frac{1}{3}$. Let us consider the least upper bound of quintessential field parameter
\begin{equation}\label{A3}
w_{q}=-\frac{1}{3}.
\end{equation}
We can compute the radius of the horizons by the solution of the quadratic equation $\Delta=0$
\begin{equation}\label{3}
r_{\pm}=\frac{M\pm\sqrt{M^{2}-\big(1-c\big)\big(a^{2}+q^{2}\big)}}{1-c},~~~~\text{where}~~~~r_{+}>r_{-}.
\end{equation}
In consideration of
\begin{equation}\label{a}
M^{2}-\big(1-c\big)\big(a^{2}+q^{2}\big)\geq0.
\end{equation}
The sum and product of the horizons are very beneficial results given as
\begin{eqnarray}
r_{+}+r_{-}&=&\frac{2M}{1-c},\label{4}  \\ r_{+}r_{-}&=&\frac{a^{2}+q^{2}}{1-c}.\label{5}
\end{eqnarray}
Presently the area of the horizons is figured out as
\begin{eqnarray}\label{2}
A_{\pm}&=&\int_{0}^{2\pi}\int_{0}^{\pi} \sqrt{g_{\theta\theta} g_{\phi\phi}}d\theta d\phi\notag\\&=&4\pi\big(r_{\pm}^{2}+a^{2}\big).
\end{eqnarray}
The angular velocity at both horizons is computed as
\begin{equation}\label{***}
\Omega_{\pm}=\frac{a}{r^{2}_{\pm}+a^{2}}.
\end{equation}
The surface gravity of both horizons is given by
\begin{eqnarray}\label{6}
\kappa_{\pm}&=&\frac{\big(1-c\big)\big(r_{\pm}-r_{\mp}\big)}{2\big(r_{\pm}^{2}+a^{2}\big)}\notag\\&=&\frac{\big(1-c\big)^{2}\big(r_{\pm}
-r_{\mp}\big)}{2\big(2Mr_{\pm}-a^{2}c-q^{2}\big)},
\end{eqnarray}
where $\kappa_{+}>\kappa_{-}$. \\The semi classical Bekenstein-Hawking \cite{24},\cite{25} entropy of the horizons is found to be
\begin{eqnarray}\label{7}
\delta_{\pm}&=&\frac{A_{\pm}}{4}\notag\\&=&\pi\bigg(\frac{2Mr_{\pm}-a^{2}c-q^{2}}{1-c}\bigg).
\end{eqnarray}
The Hawking temperature at the horizons is given as
\begin{eqnarray}\label{8}
T_{\pm}&=&\frac{\kappa_{\pm}}{2\pi}\notag\\&=&\frac{\big(1-c\big)\big(r_{\pm}-r_{\mp}\big)}{4\pi\big(r_{\pm}^{2}+a^{2}\big)}.
\end{eqnarray}
Clearly $T_{+}>T_{-}$. \\The area sum of both the horizons
\begin{equation}\label{8a}
A_{+}+A_{-}=8\pi\bigg(\frac{2M^{2}}{(1-c)^{2}}-\frac{a^{2}c+q^{2}}{1-c}\bigg).
\end{equation}
It implies that the area sum of both the horizons relies upon the mass $M$, charge $q$, spin parameter $a$ and the normalization factor $c$. Thus, the area sum is not universal. \\Also, the product of area of both the horizons is
\begin{equation}\label{9a}
A_{+}A_{-}=(8\pi)^{2}\Bigg(\frac{4J^{2}+\big(a^{2}c+q^{2}\big)^{2}}{4(1-c)^{2}}\Bigg).
\end{equation}
Now the area product is free from mass $M$ but it surely relies upon the angular momentum $J$, charge $q$, spin parameter $a$ and the normalization factor $c$. Hence, the area product is universal. \\Likewise, we have worked out at the entropy sum and product of both the horizons
\begin{equation}\label{10}
\delta_{+}+\delta_{-}=2\pi\bigg(\frac{2M^{2}}{(1-c)^{2}}-\frac{a^{2}c+q^{2}}{1-c}\bigg),
\end{equation}
and
\begin{equation}\label{11}
\delta_{+}\delta_{-}=(2\pi)^{2}\bigg(\frac{4J^{2}+\big(a^{2}c+q^{2}\big)^{2}}{4(1-c)^{2}}\bigg).
\end{equation}
Here entropy sum relies upon the mass $M$ but entropy product does not rely upon it. \\We also calculate the entropy minus
\begin{equation}\label{12}
\delta_{\pm}-\delta_{\mp}=\frac{2\pi M\big(r_{\pm}-r_{\mp}\big)}{1-c}.
\end{equation}
The sum of entropy inverse is given as
\begin{equation}\label{13}
\frac{1}{\delta_{+}}+\frac{1}{\delta_{-}}=\frac{2M^{2}-\big(1-c\big)\big(a^{2}c+q^{2}\big)}{2\pi\big[4J^{2}+\big(a^{2}c+q^{2}\big)^{2}\big]}.
\end{equation}
The minus of entropy inverse is calculated as
\begin{equation}\label{14}
\frac{1}{\delta_{\pm}}-\frac{1}{\delta_{\mp}}=\mp\frac{4M\sqrt{M^{2}-\big(1-c\big)\big(a^{2}+q^{2}\big)}}{\pi\big[4J^{2}+\big(a^{2}c+q^{2}\big)^{2}\big]}.
\end{equation}
It shows that entropy minus, sum of entropy inverse and minus of entropy inverse rely upon mass $M$. \\We also computed Kerr like bound for Kerr-Newman black hole in presence of quintessence by using Eq. (\ref{a})
\begin{equation}\label{21}
M^{4}-\big(1-c\big)\big(J^{2}+M^{2}q^{2}\big)\geq0.
\end{equation}
Also
\begin{equation}\label{22}
M^{2}=\frac{q^{2}(1-c)+\sqrt{q^{4}(1-c)^{2}+4J^{2}(1-c) }}{2}.
\end{equation}
In view of $\delta_+\geq\delta_-\geq0$, the entropy product (\ref{11}) permits
\begin{equation}\label{17}
\delta_{+}\geq\sqrt{\delta_{+}\delta_{-}}=\sqrt{(2\pi)^{2}\bigg(\frac{4J^{2}+\big(a^{2}c+q^{2}\big)^{2}}{4(1-c)^{2}}\bigg)}\geq\delta_{-},
\end{equation}
and the entropy sum (\ref{10}) allows
\begin{eqnarray*}\label{18a}
2\pi\bigg(\frac{2M^{2}}{(1-c)^{2}}-\frac{a^{2}c+q^{2}}{1-c}\bigg)=\delta_{+}+\delta_{-}\geq\delta_{+}\geq\frac{\delta_{+}\delta_{-}}{2}
=\pi\bigg(\frac{2M^{2}}{(1-c)^{2}}-\frac{a^{2}c+q^{2}}{1-c}\bigg)\geq\delta_{-}.
\end{eqnarray*}
Consequently, the entropy bound of outer horizon is figured out as
\begin{eqnarray}\label{19a}
\pi\bigg(\frac{2M^{2}}{(1-c)^{2}}-\frac{a^{2}c+q^{2}}{1-c}\bigg)\leq\delta_{+}\leq2\pi\bigg(\frac{2M^{2}}{(1-c)^{2}}-\frac{a^{2}c+q^{2}}{1-c}\bigg),
\end{eqnarray}
and the entropy bound of inner horizon is given as
\begin{equation}\label{19}
0\leq\delta_{-}\leq\sqrt{(2\pi)^{2}\bigg(\frac{4J^{2}+\big(a^{2}c+q^{2}\big)^{2}}{4(1-c)^{2}}\bigg)}.
\end{equation}
The entropy bound helps us to find the area bound which we will discover in latter section. Observe that for the case $q=0$, we attain the entropy bound for Kerr black hole in presence of quintessence. Also for $a=0$, we acquire the entropy bound for Reissner-Nordstr\"{o}m (RN) black hole in presence of quintessence.

In Table \ref{Atab:1}, we have figured out miscellaneous thermodynamical parameters for Kerr, RN and KN black holes in presence of quintessence for the quintessence field parameter $\omega{q}=-\frac{1}{3}$.
\section{Mass Formula}\label{Sec:A3}
Let us first calculate surface area to find mass formula for KN Black Hole in presence of quintessence. As reported by the Hawking's \cite{25}, the surface area of a black hole invariably non-negative, i.e.,
\begin{equation}\label{20}
dA\geq 0.
\end{equation}
The surface area of KN black hole in presence of quintessence is fixed for inner and outer horizons. It can be written as
\begin{equation}\label{20}
A_{\pm}=4\pi\Bigg(\frac{2M^{2}}{1-c}-\frac{a^{2}c+q^{2}}{1-c}\pm\frac{2\sqrt{M^{4}-\big(1-c\big)\big(J^{2}+M^{2}q^{2}\big)}}{(1-c)^{2}}\Bigg).
\end{equation}
By flipping the Eq. \ref{20}), we can compute the black hole mass or ADS mass with regard to the area of both the horizons,
\begin{equation}\label{24}
M^{2}=\frac{(1-c)^{2}}{A_{\pm}}\Bigg[\frac{A_{\pm}^{2}}{16\pi}+\frac{A_{\pm}\big(a^{2}c+q^{2}\big)}{2(1-c)}+\frac{4\pi J^{2}}{(1-c)^{2}}+\pi\bigg(\frac{a^{2}c+q^{2}}{1-c}\bigg)^{2}\Bigg].
\end{equation}
It is noteworthy that the mass can be shown with regard to the area of the outer and inner horizons. Presently, we can conclude the differential of mass for KN spacetime in presence of quintessence. It can be represented in four consistent quantities of both the horizons as
\begin{equation}\label{29}
dM=\Gamma_{\pm} dA_{\pm}+\Omega_{\pm} dJ+\Phi_{\pm}^{q}dq+\Phi_{\pm}^{a}da,
\end{equation}
where
\begin{eqnarray}
\Gamma_{\pm}&=&\frac{1}{2MA_{\pm}^{2}}\bigg[\frac{(1-c)A_{\pm}^{2}}{16\pi}-\frac{ a^{2}c(1-c)A_{\pm}}{2}-4\pi J^{2}-\pi\big(a^{2}c+q^{2}\big)^{2}\bigg],\label{A29}\\
\Omega_{\pm}&=&\frac{4\pi J}{MA_{\pm}},\label{B29}\\
\Phi_{\pm}^{q}&=&\frac{qA_{\pm}+4\pi q\big(a^{2}c+q^{2}\big)}{2MA_{\pm}},\label{C29}\\
\Phi_{\pm}^{a}&=&\frac{ac(1-c)A_{\pm}^{2}+4\pi acA_{\pm}\big(a^{2}c+q^{2}\big)}{2MA_{\pm}^{2}},\label{D29}
\end{eqnarray}
where
\begin{eqnarray}
\Gamma_{\pm}&=&\text{Effective surface tension of both the horizons},\notag\\
\Omega_{\pm}&=&\text{Angular velocity of both the horizons},\notag\\
\Phi_{\pm}^{q}&=&\text{Electromagnetic potentials of both the horizons for electric charge},\label{E29}\\
\Phi_{\pm}^{a}&=&\text{Electromagnetic potentials of both the horizons for spin parameter}.\notag
\end{eqnarray}
The mass $M$ (\ref{24}) is a homogeneous function of degree $\frac{1}{2}$ in $(A_{\pm}, J, Q^{2}, a^{2})$ i.e., $M\Big(\frac{1}{2}A_{\pm}, \frac{1}{2}J, \frac{1}{2}Q^{2}, \frac{1}{2}a^{2}\Big)=\frac{1}{2}M(A_{\pm}, J, Q^{2}, a^{2})$. By applying Euler's theorem, we can indicate the mass with regard to $ A_{\pm}$, $J$, $Q$ and $a$ as follows
\begin{equation}\label{31}
M=2\Gamma_{\pm} A_{\pm}+2\Omega_{\pm} J+\Phi_{\pm}^{q} Q+\Phi_{\pm}^{a}a,
\end{equation}
where $A_{\pm}$, $\Gamma_{\pm}$, $\Omega_{\pm}$, $\Phi_{\pm}^{q}$ and $\Phi_{\pm}^{a}$ are represented by \ref{20}), \ref{A29}), \ref{B29}), \ref{C29}) and \ref{D29}), respectively.
\section{Christodoulou's Irreducible Mass bound}\label{Sec:A4}
According to Penrose mechanism (later called Penrose process) \cite{26}, the surface area of mass increases when black hole transforms. Later on, Christodoulou has proved that irreducible mass $M_{\text{irr}}$ can not be altered by mode of any process. He showed that there is a connection between irreducible mass and surface area. Consequently, the irreducible mass $M_{\text{irr}}$ for the outer and inner horizons can be expressed as
\begin{equation}\label{30}
M_{\text{irr},\pm}=\sqrt{\frac{A_{\pm}}{16\pi}}=\frac{\sqrt{r_{\pm}^{2}+a^{2}}}{2}.
\end{equation}
We can indicate the area and angular velocity with regard to the irreducible mass
\begin{equation}\label{32}
A_{\pm}=16\pi (M_{\text{irr},\pm})^{2},
\end{equation}
also
\begin{equation}\label{33}
\Omega_{\pm}=\frac{a}{4(M_{\text{irr},\pm})^{2}}.
\end{equation}
The product of the irreducible mass $M_{\text{irr}}$ involves the product of the area
\begin{eqnarray}
M_{\text{irr},+}M_{\text{irr},-}&=&\sqrt{\frac{A_{+}A_{-}}{(16\pi)^{2}}}.\label{34}\\
&=&\frac{\sqrt{4J^{2}+\big(a^{2}c+q^{2}\big)^{2}}}{4(1-c)}.\label{34}
\end{eqnarray}
It indicates that irreducible mass product is universal because it does not rely upon the mass of the black hole.\\
The Christodoulou-Ruffini mass prescription for KN spacetime in presence of quintessence can be expressed with regard to the irreducible mass $M_{\text{irr}}$, angular momentum $J$, charge $q$, spin parameter $a$ and normalization factor $c$ as
\begin{equation}\label{36}
M^{2}=\bigg(M_{\text{irr},\pm}(1-c)+\frac{q^{2}}{4M_{\text{irr},\pm}}\bigg)^{2}+\frac{J^{2}}{4(M_{\text{irr},\pm})^{2}}+\frac{a^{2}c}{4}\bigg(2(1-c)
+\frac{a^{2}c}{4(M_{\text{irr},\pm})^{2}}+\frac{q^{2}}{2(M_{\text{irr},\pm})^{2}}\bigg).
\end{equation}
When we make charge parameter $q$ equals to zero, we acquire the following mass prescription
\begin{equation}\label{37}
M^{2}=\big(M_{\text{irr},\pm}(1-c)\big)^{2}+\frac{J^{2}}{4(M_{\text{irr},\pm})^{2}}+\frac{a^{2}c}{4}\bigg(2(1-c)
+\frac{a^{2}c}{4(M_{\text{irr},\pm})^{2}}\bigg).
\end{equation}
It is the Christodoulou-Ruffini mass prescription for Kerr spacetime in presence of quintessence. When the spin parameter $a$ becomes zero, we attain the mass formula for RN black hole in presence of quintessence.\\
Forthwith, we are interested to find area bound and irreducible mass bound. In view of $r_{+}>r_{-}$, $A_+\geq A_-\geq 0$. Formerly, the area product shows
\begin{equation}\label{38}
A_{+}\geq\sqrt{A_{+}A_{-}}=\sqrt{(8\pi)^{2}\Bigg(\frac{4J^{2}+\big(a^{2}c+q^{2}\big)^{2}}{4(1-c)^{2}}\Bigg)}\geq A_{-}.
\end{equation}
The area sum indicates
\begin{eqnarray}\label{39}
8\pi\bigg(\frac{2M^{2}}{(1-c)^{2}}-\frac{a^{2}c+q^{2}}{1-c}\bigg)=A_{+}+A_{-}\geq A_{+}\geq\frac{A_{+}+A_{-}}{2}=4\pi\bigg(\frac{2M^{2}}{(1-c)^{2}}-\frac{a^{2}c+q^{2}}{1-c}\bigg).
\end{eqnarray}
Therefore, the area bound for the outer horizon is given by
\begin{equation}\label{40}
4\pi\bigg(\frac{2M^{2}}{(1-c)^{2}}-\frac{a^{2}c+q^{2}}{1-c}\bigg)\leq A_{+} \leq 8\pi\bigg(\frac{2M^{2}}{(1-c)^{2}}-\frac{a^{2}c+q^{2}}{1-c}\bigg).
\end{equation}
The area bound for the inner horizon is given by
\begin{equation}\label{41}
0\leq A_{-}\leq \sqrt{(8\pi)^{2}\Bigg(\frac{4J^{2}+\big(a^{2}c+q^{2}\big)^{2}}{4(1-c)^{2}}\Bigg)}.
\end{equation}
\begin{table}[h!]
\caption{A comparison of thermodynamical parameters for Kerr, RN and KN
black holes in the presence of quintessence for the quintessence field paramter $\omega_{q}=-\frac{1}{3}$.} \label{Atab:1}
{\renewcommand{\arraystretch}{1.55}
\begin{tabular}{|C{3.25cm}|C{4cm}|C{4cm}|C{4.8cm}|}
 \hline
$\text{Parameter}$ & $\text{Kerr BH}$ & $\text{RN BH}$ & $\text{KN BH}$ \\
\hline

$r_{\pm}$  & $\frac{M\pm\sqrt{M^{2}-a^{2}(1-c)}}{1-c}$ & $\frac{M\pm\sqrt{M^{2}-q^{2}(1-c)}}{1-c}$ & $\frac{M\pm\sqrt{M^{2}-\big(1-c\big)\big(a^{2}+q^{2}\big)}}{1-c}$    \\
$r_{+}+r_{-}$ & $\frac{2M}{1-c}$ & $\frac{2M}{1-c}$ & $\frac{2M}{1-c}$   \\
$r_{+}r_{-}$ & $\frac{a^{2}}{1-c}$ & $\frac{q^{2}}{1-c}$ & $\frac{a^{2}+q^{2}}{1-c}$   \\

$A_{\pm}$  & $4\pi\Big(\frac{2Mr_{\pm}-a^{2}c}{1-c}\Big)$ & $4\pi\Big(\frac{2Mr_{\pm}-q^{2}}{1-c}\Big)$ & $4\pi\Big(\frac{2Mr_{\pm}-a^{2}c-q^{2}}{1-c}\Big)$    \\
$A_{+}+A_{-}$ & $8\pi\Big(\frac{2M^{2}}{(1-c)^{2}}-\frac{a^{2}c}{1-c}\Big)$ & $8\pi\Big(\frac{2M^{2}}{(1-c)^{2}}-\frac{q^{2}}{1-c}\Big)$ & $8\pi\Big(\frac{2M^{2}}{(1-c)^{2}}-\frac{a^{2}c+q^{2}}{1-c}\Big)$   \\
$A_{+}A_{-}$ & $(8\pi)^{2}\Big(\frac{4J^{2}+a^{4}c^{4}}{4(1-c)^{2}}\Big)$ & $(8\pi)^{2}\bigg(\frac{4J^{2}+q^{4}}{4(1-c)^{2}}\bigg)$ & $(8\pi)^{2}\bigg(\frac{4J^{2}+\big(a^{2}c+q^{2}\big)^{2}}{4(1-c)^{2}}\bigg)$   \\

$\delta_{\pm}$  & $\pi\Big(\frac{2Mr_{\pm}-a^{2}c}{1-c}\Big)$  & $\pi\Big(\frac{2Mr_{\pm}-q^{2}}{1-c}\Big)$ & $\pi\Big(\frac{2Mr_{\pm}-a^{2}c-q^{2}}{1-c}\Big)$    \\
$\delta_{+}+\delta_{-}$ & $2\pi\Big(\frac{2M^{2}}{(1-c)^{2}}-\frac{a^{2}c}{1-c}\Big)$ & $2\pi\Big(\frac{2M^{2}}{(1-c)^{2}}-\frac{q^{2}}{1-c}\Big)$ & $2\pi\Big(\frac{2M^{2}}{(1-c)^{2}}-\frac{a^{2}c+q^{2}}{1-c}\Big)$   \\
$\delta_{+}\delta_{-}$ & $(2\pi)^{2}\Big(\frac{4J^{2}+a^{4}c^{4}}{4(1-c)^{2}}\Big)$ & $(2\pi)^{2}\bigg(\frac{4J^{2}+q^{4}}{4(1-c)^{2}}\bigg)$ & $(2\pi)^{2}\bigg(\frac{4J^{2}+\big(a^{2}c+q^{2}\big)^{2}}{4(1-c)^{2}}\bigg)$   \\

$\kappa_{\pm}$  & $\frac{(1-c)^{2}(r_{\pm}-r_{\mp})}{2(2Mr_{\pm}-a^{2}c)}$ & $\frac{(1-c)^{2}(r_{\pm}-r_{\mp})}{2(2Mr_{\pm}-q^{2})}$ & $\frac{(1-c)^{2}(r_{\pm}-r_{\mp})}{2(2Mr_{\pm}-a^{2}c-q^{2})}$    \\
$\kappa_{+}+\kappa_{-}$ & $\frac{4M\big[a^{2}(1-c)-M^{2}\big]}{4J^{2}+a^{4}c^{2}}$ & $\frac{4M\big[q^{2}(1-c)-M^{2}\big]}{4J^{2}+q^{4}}$ & $\frac{4M\big[\big(1-c\big)\big(a^{2}+q^{2}\big)-M^{2}\big]}{4J^{2}+\big(a^{2}c+q^{2}\big)^{2}}$   \\
$\kappa_{+}\kappa_{-}$ & $\frac{(1-c)^{2}\big[a^{2}(1-c)-M^{2}\big]}{4J^{2}+a^{4}c^{2}}$ & $\frac{(1-c)^{2}\big[q^{2}(1-c)-M^{2}\big]}{4J^{2}+q^{4}}$ & $\frac{(1-c)^{2}\big[\big(1-c\big)\big(a^{2}+q^{2}\big)-M^{2}\big]}{4J^{2}+\big(a^{2}c+q^{2}\big)^{2}}$   \\

$T_{\pm}$  & $\frac{(1-c)^{2}(r_{\pm}-r_{\mp})}{4\pi(2Mr_{\pm}-a^{2}c)}$ & $\frac{(1-c)^{2}(r_{\pm}-r_{\mp})}{4\pi(2Mr_{\pm}-q^{2})}$ & $\frac{(1-c)^{2}(r_{\pm}-r_{\mp})}{4\pi(2Mr_{\pm}-a^{2}c-q^{2})}$    \\
$T_{+}+T_{-}$ & $\frac{2M\big[a^{2}(1-c)-M^{2}\big]}{\pi\big[4J^{2}+a^{4}c^{2}\big]}$ & $\frac{2M\big[q^{2}(1-c)-M^{2}\big]}{\pi\big[4J^{2}+q^{4}\big]}$ & $\frac{2M\big[\big(1-c\big)\big(a^{2}+q^{2}\big)-M^{2}\big]}{\pi\big[4J^{2}+\big(a^{2}c+q^{2}\big)^{2}\big]}$   \\
$T_{+}T_{-}$ & $\frac{(1-c)^{2}\big[a^{2}(1-c)-M^{2}\big]}{(2\pi)^{2}\big[4J^{2}+a^{4}c^{2}\big]}$ & $\frac{(1-c)^{2}\big[q^{2}(1-c)-M^{2}\big]}{(2\pi)^{2}\big[4J^{2}+q^{4}\big]}$ & $\frac{(1-c)^{2}\big[\big(1-c\big)\big(a^{2}+q^{2}\big)-M^{2}\big]}{(2\pi)^{2}\big[4J^{2}+\big(a^{2}c+q^{2}\big)^{2}\big]}$   \\

$M_{\text{irr},\pm}$  & $\sqrt{\frac{2Mr_{\pm}-a^{2}c}{4(1-c)}}$ & $\sqrt{\frac{2Mr_{\pm}-q^{2}}{4(1-c)}}$ & $\sqrt{\frac{2Mr_{\pm}-a^{2}c-q^{2}}{4(1-c)}}$    \\
$\big(M_{\text{irr},+}\big)^{2}+\big(M_{\text{irr},-}\big)^{2}$ & $\frac{2M^{2}-a^{2}c}{2(1-c)}$ & $\frac{2M^{2}-q^{2}}{2(1-c)}$ & $\frac{2M^{2}-a^{2}c-q^{2}}{2(1-c)}$   \\
$M_{\text{irr},+}M_{\text{irr},-}$ & $\frac{\sqrt{4J^{2}+a^{4}c^{4}}}{4(1-c)}$ & $\frac{\sqrt{4J^{2}+q^{4}}}{4(1-c)}$ & $\frac{\sqrt{4J^{2}+\big(a^{2}c+q^{2}\big)^{2}}}{4(1-c)}$   \\

$\Omega_{\pm}$  & $\frac{a(1-c)}{2Mr_{\pm}-a^{2}c}$ & $0$ & $\frac{a(1-c)}{2Mr_{\pm}-a^{2}c-q^{2}}$    \\
$\Omega_{+}+\Omega_{-}$ & $\frac{a\big[2M^{2}-a^{2}c(1-c)\big]}{2\big[4J^{2}+a^{4}c^{2}\big]}$  & $0$ & $\frac{a\big[2M^{2}-\big(1-c\big)\big(a^{2}c+q^{2}\big)\big]}{2\big[4J^{2}+\big(a^{2}c+q^{2}\big)^{2}\big]}$   \\
$\Omega_{+}\Omega_{-}$ & $\frac{a^{2}(1-c)^{2}}{4J^{2}+a^{4}c^{2}}$  & $0$ & $\frac{a^{2}(1-c)^{2}}{4J^{2}+\big(a^{2}c+q^{2}\big)^{2}}$   \\ \hline

\end{tabular}
}
\end{table}

We obtain the irreducible mass bound for the outer horizon of KN black hole in quintessence from the inequality \ref{40})
\begin{equation}\label{42}
\frac{\sqrt{\frac{2M^{2}}{(1-c)^{2}}-\frac{a^{2}c+q^{2}}{1-c}}}{2}\leq M_{\text{irr},+} \leq \frac{\sqrt{\frac{2M^{2}}{(1-c)^{2}}-\frac{a^{2}c+q^{2}}{1-c}}}{\sqrt{2}},
\end{equation}
and for the inner horizon, we have
\begin{equation}\label{43}
0\leq M_{\text{irr},-}\leq\frac{\big[4J^{2}+\big(a^{2}c+q^{2}\big)^{2}\big]^{\frac{1}{4}}}{2\sqrt{1-c}}.
\end{equation}
The irreducible mass bound \ref{42}) is called Penrose inequality \cite{23}.
\section{Conclusion}\label{Sec:A5}
In this research work, predominantly we acquired area, angular velocity, entropy, surface gravity and Hawking temperature of inner and outer horizons for Kerr-Newman black hole in presence of quintessence. Additionally, we examined area sum, area product, entropy sum and entropy product. We investigated that the area product and entropy product are free from mass $M$ but these products surly rely upon the angular momentum $J$, charge $q$, spin parameter $a$ and the normalization factor $c$. We observed that these thermodynamic products are universal. We have studied that the area sum and entropy sum rely upon the mass $M$, charge $q$, spin parameter $a$ and the normalization factor $c$, so these sums are not universal. The black hole mass and Christodoulou-Ruffini mass for Kerr-Newman black hole in quintessence have been found. We have extracted the entropy bound from the area bound. We have studied that the irreducible mass product does not rely upon the mass $M$, so this product is also universal. We have indicated the area, angular velocity and Christodoulou-Ruffini mass in terms of irreducible mass. We have derived the Penrose inequality and have discussed the microscopic nature of the entropy.

\end{document}